\documentstyle[12pt]{article}
\newcommand {\dc}{\cal D}
\newcommand {\ps}{\sf p}

\newcommand \corr[1]{\left\langle{#1}\right\rangle}

\newcommand {\pprod}{\mathop{{\prod}'}}
\newcommand {\pw}{\displaystyle \mathop{{\prod^{d+2}_{i=1}}{{w_i}^{w_i}}}}
\newcommand {\kw}{\displaystyle \mathop{{\prod^{d+2}_{i=1}}{{w_i}!}}}

\newcommand {\Kappa}{\mbox{\Large $\kappa$}}
\newcommand {\Eta}{\mbox{\large $\eta$}}

\newcommand {\beq}{\begin{eqnarray}}
\newcommand {\eeq}{\end{eqnarray}}
\newcommand {\beqs}{\begin{eqnarray*}}
\newcommand {\eeqs}{\end{eqnarray*}}
\newcommand {\del}{\partial}

\newcommand \cm[2]{\left[ {#1}\, ,{#2} \right]}
\newcommand \ac[2]{\left\{ {#1}\, ,{#2}\right\}}
\newcommand \th[1]{{\Theta}_{#1}}

\newcommand {\bc}{\bf C}

\newcommand {\vpi}{\varpi}

\newcommand {\Om}{\Omega}
\newcommand \hm[2]{\widehat{M}^{({#1})}_{#2}}
\newcommand \om[1]{{\omega}_{#1}}

\newcommand {\amb}{{\bf P}_{d+1} [2,2,2,\cdots ,2,2,1,1](2(d+1))}

\newcommand {\kae}{\mbox{K{\"a}hler}}

\newcommand{\pr}{\hspace{\parindent}}

\begin{document}
\setlength{\oddsidemargin}{0cm}
\setlength{\baselineskip}{6.3mm}

\begin{titlepage}
    \begin{normalsize}
     \begin{flushright}
            { YITP/U-95-13} \\
            % {preliminary version} \\
            % April 1995
     \end{flushright}
    \end{normalsize}
    \begin{LARGE}
       \vspace{1cm}
       \begin{center}
        { Three Point Functions on the Sphere\\
            of Calabi-Yau d-Folds\\
                  }
       \end{center}
    \end{LARGE}

   \vspace{5mm}

\begin{center}
           Katsuyuki S{\sc ugiyama}\footnote{E-mail address:
             ksugi@yisun1.yukawa.kyoto-u.ac.jp} \\
       \vspace{4mm}
                  {\it Uji Research Center, } \\
                  {\it Yukawa Institute for Theoretical Physics, } \\
                  {\it Kyoto University,
                  Uji 611, Japan} \\
       \vspace{1cm}

     \begin{large} ABSTRACT \end{large}
\par
  \end{center}
\begin{quote}
 \begin{normalsize}

Using mirror symmetry in Calabi-Yau manifolds M, three point functions
of A(M)-model operators on the genus $0$ Riemann surface in cases of
one-parameter families of $d$-folds
realized as Fermat type
hypersurfaces embedded in weighted projective spaces and a
two-parameter family of $d$-fold embedded in a weighted projective
space ${\amb}$
%${{\bf P}_{d+1}
%[{{\underbrace{2,2,\cdots ,2}_{d\,\,times}},1,1}](2(d+1))}$
are studied. These three point functions
%${{\left({\Kappa^{(l-1)}_{a}}\right)}_{bc}}$
%=
${\corr{\,{{\cal O}^{(1)}_{a}}\,
{{\cal O}^{(l-1)}_{b}}\,
{{\cal O}^{(d-l)}_{c}}\,
}}$ are expanded by indeterminates ${q_l}$=${e^{2\pi i {t_l}}}$
associated with a set of {\kae} coordinates $\{{t_l}\}$ and their
expansion coefficients count the number of maps with a definite
degree which map each of three points ${0,1,\mbox{\,and\,}\infty }$
on the world sheet on some homology cycle of M associated with a
cohomology element. From these analyses,
we can read fusion structure of Calabi-Yau
A(M)-model operators. In our cases they constitute a subring of a
total quantum cohomology ring of the A(M)-model operators. In fact we
switch off all perturbation operators on the topological theories
except for marginal ones associated with {\kae} forms of M. For that
reason, the charge conservation of operators turns out to be a
classical one. Furthermore because their
first Chern classes ${c_1}$ vanish, their topological selection rules
do not depend on the degree of maps, (especially a nilpotent
property of operators ${
{{\cal O}^{(1)}}{{\cal O}^{(d)}}=0
}$ is satisfied). Then these fusion couplings $\{{\Kappa_l}\}$ are
represented as some series adding up all degrees of maps.

 \end{normalsize}
\end{quote}

\end{titlepage}
\vfill

\section{Introduction}

\pr
When one considers correlation functions of an N=2 non-linear sigma model
with a Calabi-Yau target space, they depend not only on the moduli space
of the Riemann surface, but also on the properties of the target Calabi-Yau
manifold (especially on the Calabi-Yau moduli spaces).
There are two quasi-topological field theories (the A-model and the B-model)
{\cite{W}} obtained by twisting the N=2 non-linear sigma model {\cite{WIT,EY}}
, which describe two distinct Calabi-Yau
moduli spaces (a {\kae} structure moduli space and
a complex structure moduli space, respectively).
As is well-known, the correlators of the B-model receive no quantum
corrections {\cite{W,DG,CDGP,AM}},
but the correlation functions of the A-model have
non-perturbative corrections which originate in holomorphic maps from the
Riemann surface to the Calabi-Yau target space.

Recently by the discovery of the mirror symmetry {\cite{GP,CLS}}
between the A(M)-model and the B(W)-model
for the mirror pairs (M, W), it is becoming possible to
obtain the A(M)-model correlators from the B(W)-model ones indirectly but
in a form involving all the quantum corrections {\cite{CDGP,Y}}.
Now it may fairly be said that the analyses of the Calabi-Yau 3-folds under
the mirror symmetries have been established {\cite{KT,F,KT2,CDFKM,CFKM,BCDFHJQ,
LSW,HKT,BV}}. In this article, we study the
A(M)-model correlators of some d-dimensional Calabi-Yau manifolds as
mathematical physics applications under the mirror symmetry
furthermore {\cite{NS,GMP,JN}}.

Because both the A-model and the B-model are pseudo-topological theories, they
are characterized by their two point functions and three point functions
 which play important roles as the constituent blocks in these models.
The two point functions of the A-model are called the topological metrics
and receive no quantum corrections {\cite{W}}.
On the other hand, the three point
functions of the A-model have non-perturbative quantum corrections and
have information about the fusion structure of the observables (the classical
cohomology structure and quantum corrections).
Because various physical quantities are determined by this fusion structure
of these operators, it is important to study the properties of the
operator products (the commutativity, associativity, and the existence of the
unit operator) or the factorization of the multi-point functions.

In this article, we take a genus zero Riemann surface as a world sheet,
(complex) d-dimensional Calabi-Yau target spaces and analyze
the properties of the three point functions
under the mirror symmetry in order to clarify the
effects of the non-perturbative instanton corrections.

\section{The Calabi-Yau d-folds}

\pr
We consider d-dimensional Calabi-Yau manifolds $M$ in two cases;
\beqs
\bullet &&\mbox{Case I}\,\,\,\mbox{(one-parameter families)}\\
&&\,\,\mbox{(I-1)}
\,\,\,M\,;\,{\ps}={X_1^{d+2}}+{X_2^{d+2}}+\cdots +{X_{d+2}^{d+2}}\\
&&\hspace{2cm}-(d+2)\psi ({X_1}{X_2}\cdots {X_{d+2}})=0
\,\,\,in \,\,\, {CP}^{d+1}\,\,\,,\\
&&\,\,\mbox{(I-2)}
\,\,\,M\,;\,{\ps}={X_1^{2}}+{X_2^{2(d+1)}}+\cdots +{X_{d+2}^{2(d+1)}}\\
&&\hspace{2cm}-2(d+1)\psi ({X_1}{X_2}\cdots {X_{d+2}})=0
\,\,\,in \,\,\,{{\bf P}_{d+1}}[{d+1,
\underbrace{1,1,\cdots ,1}_{(d+1)\,\,times}}](2(d+1))\,\,\,,\\
&&\,\,\mbox{(I-0)}
\,\,\,M\,;\,{\ps}={X_1^{l_1}}+{X_2^{l_2}}+\cdots +{X_{d+2}^{l_{d+2}}}\\
&&\hspace{2cm}-D\psi ({X_1}{X_2}\cdots {X_{d+2}})=0
\,\,\,in \,\,\, {{\bf P}_{d+1}}[{{w_1},{w_2},\cdots ,
{w_{d+2}}}](D)\,\,\,,\\
&&\hspace{3cm} D:={\sum^{d+2}_{i=1}}{w_i}\,\,,\,\,
{l_i}:=\frac{D}{w_i}\,\,,\,\,
{w_{d+2}}:=1\,\,\,,\\
\bullet &&\mbox{Case II}\,\,\,\mbox{(two-parameter family)}\\
&&\,\,\mbox{(II)}
\,\,\,M\,;\,{\ps}={X_1^{d+1}}+{X_2^{d+1}}+\cdots +{X_{d}^{d+1}}+
{X_{d+1}^{2(d+1)}}+{X_{d+2}^{2(d+1)}}-2\phi
{{({X_{d+1}}{X_{d+2}})}^{d+1}}\\
&&\hspace{1.2cm}-2(d+1)\psi ({X_1}{X_2}\cdots {X_{d+2}})
=0
\,\,\,
%&&\hspace{3cm}
in \,\,\,
{{\bf P}_{d+1}
[{{\underbrace{2,2,\cdots ,2}_{d\,\,times}},1,1}](2(d+1))}\,\,\,.
\eeqs
The parameters ${\psi}$ and ${\phi}$ control the deformation of the
complex structures. In case (I-0), the sets of integers
$({{l_1},{l_2},\cdots ,{l_{d+2}}})$
are given in some lower dimensional cases;
\beqs
%\begin{array}{ll}
&& \bullet \,\,\,d=3\\
&& \hspace{2.5cm}
\mbox{
\begin{tabular}{|c|ccccc|} \hline
{d=3} & ${l_1}$ & ${l_2}$ & ${l_3}$ & ${l_4}$ & ${l_5}$ \\ \hline
(1) & 5 & 5 & 5 & 5 & 5 \\ \hline
(2) & 3 & 6 & 6 & 6 & 6 \\ \hline
(3) & 2 & 8 & 8 & 8 & 8 \\ \hline
(4) & 2 & 5 & 10 & 10 & 10 \\ \hline
\end{tabular}
}\\
&& \bullet \,\,\,d=4\\
&& \hspace{2.5cm}
\mbox{
\begin{tabular}{|c|cccccc|} \hline
{d=4} & ${l_1}$ & ${l_2}$ & ${l_3}$ & ${l_4}$ & ${l_5}$ & ${l_6}$ \\ \hline
(1) & 2 & 10 & 10 & 10 & 10 & 10 \\ \hline
(2) & 6 & 6 & 6 & 6 & 6 & 6  \\ \hline
\end{tabular}
}\\
&& \bullet \,\,\,d=5\\
&& \hspace{2.5cm}
\mbox{
\begin{tabular}{|c|ccccccc|} \hline
{d=5} & ${l_1}$ & ${l_2}$ & ${l_3}$ & ${l_4}$ & ${l_5}$ & ${l_6}$ &
 ${l_7}$ \\ \hline
(1) & 4 & 8 & 8 & 8 & 8 & 8 & 8 \\ \hline
(2) & 3 & 9 & 9 & 9 & 9 & 9 & 9 \\ \hline
(3) & 2 & 12 & 12 & 12 & 12 & 12 & 12 \\ \hline
(4) & 3 & 4 & 12 & 12 & 12 & 12 & 12 \\ \hline
(5) & 2 & 7 & 14 & 14 & 14 & 14 & 14 \\ \hline
(6) & 7 & 7 & 7 & 7 & 7 & 7 & 7 \\ \hline
\end{tabular}
}
%\end{array}
\eeqs
Because the case (I-0) contains cases (I-1) and (I-2), we analyze cases
(I-0) and (II) in the following.
Mirror partners $W$ of these Fermat type Calabi-Yau d-folds $M$
((I-0), (II)) are given as orbifolds
divided by some maximally invariant discrete groups $G$,
\[
W\,;\,\{{p=0}\}/G \,\,\,.
\]
When one thinks about the Hodge structure of the $G$-invariant part of the
cohomology group ${H^d}(W)$, Hodge numbers of $W$ are written;
\beqs
&&\mbox{(I-0)}\,;\,{h^{d,0}}={h^{d-1,1}}=\cdots ={h^{1,d-1}}
={h^{0,d}}=1\,\,\,,\\
&&\mbox{(II)}\,;\,{h^{d,0}}={h^{0,d}}=1\,\,,\,\,{h^{d-1,1}}={h^{d-2,2}}=\cdots
={h^{2,d-2}}={h^{1,d-1}}=2\,\,\,.
\eeqs

\section{The One-Parameter Families}

\pr
Firstly we investigate the case (I-0).
The deformation of the complex structure of $W$ is controlled by the structure
of the Hodge decomposition of ${H^d}(W)$ and the information of the
decomposition is given by the period matrix $P$ of $W$.
The period matrix is defined by using homology d-cycles
${\gamma_j}\in {H_d}(W)$ and cohomology elements ${\alpha_i}\in
{{\cal F}^{d-i}}={H^{d,0}}\oplus {H^{d-1,1}}\oplus \cdots \oplus {H^{d-i,i}}$
and its matrix elements ${P_{ij}}\,\,(0\leq i \leq d\,,\,0\leq j \leq d)$
are expressed as,
\[
{P_{ij}}:={\int_{\gamma_j}}{\alpha_i}\,\,\,.
\]
Especially the ${\alpha_0}={\Omega}$ is a globally defined
nowhere-vanishing holomorphic
d-form of $W$ and can be expressed for the Fermat type
hypersurface ${\ps}$ by {\cite{LSW,G,BG}},
\beqs
&&\Om :={\int_{\gamma}}{\frac{d\mu}{\ps}}\,\,\,,\\
&&d\mu :={\sum^{d+2}_{a=1}}{{(-1)}^{a-1}}{w_a}\,
{X_a} d{X_1}\wedge d{X_2}\wedge \cdots \wedge
\widehat{d{X_a}} \wedge \cdots \wedge d{X_{d+2}}\,\,\,,
\eeqs
where ${\gamma}$ is a small one-dimensional cycle
winding around the hypersurface defined as a zero locus of ${\ps}$,
\[
{Z_p}:=\{{\left({X_1},{X_2},\cdots ,{X_{d+2}}\right)\,;\,{\ps}=0
}\}\,\,\,.
\]
Also the {${\alpha_i}$}'s are defined as,
\beqs
&& {\alpha_i}:={\Theta^{i}_{z}}\Om \,\,,\,\,{\Theta_z}:=z\frac{d}{dz}\,\,\,,\\
&& z:={{({D\psi})}^{-D}}\,\,\,.
\eeqs
Now we pick the elements of the period matrix $P$ in the zero-th row,
\[
{P_{0j}}={\int_{\gamma_j}}\Om \equiv {\varpi_j}\,\,\,.
\]
Using the explicit formula of the d-form $\Om $, we obtain a differential
equation satisfied by {${\varpi_j}$}'s,
%%%%%%%%%%%%%OB%%%%%%%%%%%%%%%%%%%%%11111111111111
\beq
&&\left[{{\th{z}}^{d+1} -\gamma \pprod^{D-1}_{l=1}
\left({\th{z}+\frac{l}{D}}\right)}\right]{\varpi_j}=0
\,\,\,,\,\,\,
\gamma :=\frac{D^D}{\pw}\,\,\,,
\label{eqn:aagkz2}
\eeq
where the product ${\pprod^{D-1}_{l=1}}$ means that the variable $l$
runs over integers ranging from one to $(D-1)$ which are not divisible by any
${l_i}$,
\[
\pprod^{D-1}_{l=1} :=
\begin{array}[t]{c}
{\displaystyle \prod^{D-1}_{l=1}} \\
{\scriptstyle \begin{array}{c}
              {\scriptstyle  {l_i} {\not{|}} {l}}
%               { \mbox{\small for some}\,\, {\scriptstyle m\in {\bz}}}
               \end{array} }
\end{array} \,\,\,.
\]
We solve this equation ({\ref{eqn:aagkz2}}) in a series form,
\beqs
&&{P_{0j}}={{\vpi}_j} ({z}):=
{\sum^{j}_{l=0}}\,\frac{1}{l!}\,
{\left({\frac{\log z}{2\pi i}}\right)}^{l}
\times {\sum^{\infty}_{m=0}}{b_{j-l,m}}\cdot {z^m}\,\,\,,\\
&&{b_{n,m}}:=\frac{1}{n!} {\left({\frac{1}{2\pi i}\cdot
\frac{\del}{\del\rho}}\right)}^{n}
\left\{{
\frac{\Gamma ({D({m+\rho})+1})}{\Gamma ({D{\rho}+1})}\cdot
{\prod^{d+2}_{i=1}}
\frac{\Gamma ({{w_i}{\rho}+1})}{\Gamma ({{w_i}({m+\rho})+1})}
}\right\} {\Biggr|_{\rho =0}} \,\,\,,
\eeqs
and construct the period matrix $P$ of $W$,
\[
{P_{ij}}={\int_{\gamma_j}}{\Theta^{i}_{z}}\Om ={\Theta^{i}_{z}}{\varpi_j}
\,\,\,.
\]
In this process, we do not construct homology d-cycles explicitly. In order to
study properties of these homology d-cycles
${\{{{\gamma_0},{\gamma_1},\cdots ,{\gamma_d}}\}}$,
we perform a monodromy transformation $T$ about the point ${z=0}$,
\[
T\,;\,z\rightarrow {e^{2\pi i}}z\,\,\,.
\]
Then the period matrix $P$ changes into a form,
\beqs
&& P\rightarrow P\cdot A\,\,,\,\,A=\exp (N)\,\,\,,\\
&& N:=\left(
\begin{array}{cccccc}
0 & 1 &  &  &  &  \\
  & 0 & 1 &  &  &   \\
 &  & \ddots  & \ddots  &  &  \\
  &   &  & 0  & 1  &  \\
  &   &  &    & 0  & 1 \\
  &   &  &    &    & 0
\end{array}
\right)
 \,\,\,,\,\,\,({{N^{d+1}}=0})\,\,\,,
\eeqs
and we can read the action of the monodromy transformation $T$ on the set of
cycles $({{\gamma_0}\,{\gamma_1}\,\cdots \,{\gamma_{d}}})$, which we use
implicitly in defining the period matrix $P$,
\[
T\,({{\gamma_0}\,{\gamma_1}\,\cdots \,{\gamma_{d}}})=
({{\gamma_0}\,{\gamma_1}\,\cdots \,{\gamma_{d}}})\,A\,\,\,.
\]
A peculiar property ${{N^{d+1}}=0}$ of the above monodromy matrix
${A=\exp (N)}$ indicates the
maximally unipotent monodromy condition of these homology cycles
at the point ${z=0}$ and is consistent with the usual mirror conjecture.
When one studies the variation of the complex structure of $W$, one deforms
cohomology elements but fixes homology cycles. So we investigate the set of
the cohomology classes $({{\alpha_0}\,,\,{\alpha_1}\,,\,\cdots
\,,\,{\alpha_{d}}})$ $=$
$({{\Om}\,,\,{\Theta_{z}}{\Om}\,,\,\cdots \,,\,{\Theta^{d}_{z}}\Om })$.
The cohomology group elements ${\alpha_l}={\Theta^{l}_{z}}\Om $ are
expressed as,
\beqs
&&{\Theta^{l}_{z}}\Om =
{\sum^{l}_{m=1}}\int \frac{\varphi_{lm}}{{\ps}^{m+1}}\,d\mu \,\,\,,\\
&&{\varphi_{lm}}=\left\{{
  {\sum^{m}_{k=1}} {{(-1)}^{m-k}}\cdot {\mbox{}_m}{C_k}\cdot
{{\left({\frac{-k}{D}}\right)}^{l}}
}\right\}\times {{({D\psi})}^{m}}\cdot
{{({X_1}{X_2}\cdots {X_{d+2}})}^{m}} \,\,\,.
\eeqs
When one deforms the complex structure of $W$, the set of cohomology classes
 is modified and the structure of the Hodge decomposition
${{\oplus^{d}_{p=0}}{H^{d-p,p}}(W)}$ undergoes a change. In order to look
into the variation of the complex structure of $W$, we change the period
matrix $P$ into an upper triangular one $\Phi $ with unit diagonal
elements by the sweeping-out method.
In this operation, the set of homology d-cycles
${({{\gamma_0}\,{\gamma_1}\,\cdots \,{\gamma_{d}}})}$ remains
unchanged, but the cohomology basis
${({{\alpha_0}\,{\alpha_1}\,\cdots \,{\alpha_{d}}})}$ turns into a new basis
${({{\tilde{\alpha}_0}\,{\tilde{\alpha}_1}\,\cdots \,{\tilde{\alpha}_{d}}})}$,
\beqs
&&{\tilde{\alpha}_0}:=\frac{1}{\varpi_0}{\alpha_0}\in {H^{d,0}}\,\,\,,\\
&&{\tilde{\alpha}_l}:=
\frac{1}{\tilde{\Kappa}_{l-1}}{\Theta_{z}}
\frac{1}{\tilde{\Kappa}_{l-2}}{\Theta_{z}}\cdots {\Theta_{z}}
\frac{1}{\tilde{\Kappa}_{1}}{\Theta_{z}}
\frac{1}{\tilde{\Kappa}_{0}}{\Theta_{z}}
\left({\frac{\alpha_0}{\varpi_0}}\right)\in {{\cal F}^{d-l}}
\,\,\,,\,\,\,({1\leq l \leq d})\,\,\,,\\
&&{\tilde{\Kappa}_0}:={\Theta_{z}}
\left({{\int_{\gamma_1}}{\tilde{\alpha}_0}}\right)=
{\Theta_{z}}{\omega_1}\,\,\,,\\
&&{\tilde{\Kappa}_m}:={\Theta_{z}}
\left({{\int_{\gamma_{m+1}}}{\tilde{\alpha}_{m}}}\right)=
{\Theta_{z}}\frac{1}{\tilde{\Kappa}_{m-1}}{\Theta_{z}}
\frac{1}{\tilde{\Kappa}_{m-2}}{\Theta_{z}}\cdots {\Theta_{z}}
\frac{1}{\tilde{\Kappa}_{1}}{\Theta_{z}}
\frac{1}{\tilde{\Kappa}_{0}}{\Theta_{z}}\,{\omega_{m+1}}
\,\,\,,\,\,\,({1\leq m \leq d-1})\,\,\,,\\
&&{\omega_n}:=\frac{\varpi_{n}}{\varpi_0}\,\,\,.
\eeqs
Using this new basis ${\{{\tilde{\alpha}_i}\}}$, we can write down
the resulting period matrix $\Phi $,
\beqs
&&{\Phi_{ij}}:={\int_{\gamma_j}}{\tilde{\alpha}_i}\,\,\,,\\
&&\Phi =\left(
\begin{array}{ccccccccc}
1 & \Phi_{01} & \Phi_{02} & \Phi_{03} &
\Phi_{04} & \cdots & \Phi_{0\,d-2} &
\Phi_{0\,d-1} & \Phi_{0\,d} \\
  & 1 & \Phi_{12} & \Phi_{13} & \Phi_{14} & \cdots &
 \Phi_{1\,d-2} & \Phi_{1\,d-1} & \Phi_{1\,d} \\
  &   & 1 & \Phi_{23} & \Phi_{24} & \cdots &
 \Phi_{2\,d-2} & \Phi_{2\,d-1} & \Phi_{2\,d} \\
  &   &  & 1 & \Phi_{34} & \cdots &
 \Phi_{3\,d-2} & \Phi_{3\,d-1} & \Phi_{3\,d} \\
  &   &  &   & \ddots & \ddots &
 \vdots & \vdots & \vdots \\
  &   &  &   &        & 1 &
\Phi_{d-3\,d-2} & \Phi_{d-3\,d-1} & \Phi_{d-3\,d} \\
  &   &  &   &        &   &
1  & \Phi_{d-2\,d-1} & \Phi_{d-2\,d} \\
  &   &  &   &        &   &
  & 1  & \Phi_{d-1\,d} \\
{\mbox{\Large $O$}}  &   &  &   &        &   &
  &   & 1
\end{array}
\right)\,\,\,.
\eeqs
To investigate the complex structure of $W$, we think about a
differential equation of $P $,
%%%%%%%%%%%%%%%%%%%%22222222222222222222222
\beqs
&&{\th{z}}\,P (z)={\widetilde{C}_z}\,P (z)\,\,\,,\\
&&{\widetilde{C}_z}:=\left(
\begin{array}{cccccc}
0 & 1 &  &  &  &  \\
  & 0 & 1 &  &  &   \\
 &  & \ddots  & \ddots  &  &  \\
  &   &  & 0  & 1  &  \\
  &   &  &    & 0  & 1 \\
{\sigma_{N-1}}  & {\sigma_{N-2}}  & {\sigma_{N-3}} & \cdots   &
{\sigma_{2}} & {\sigma_{1}}
\end{array}
\right) \,\,\,,\\
&&{\sigma_{m}}:={\displaystyle \frac{\gamma z}{1-\gamma z}}
\begin{array}[t]{c}
{\displaystyle {\mathop{{\sum}'}}}\\
{\scriptstyle 1\leq {n_1} < {n_2} < \cdots < {n_m} \leq D-1}
\end{array}
\frac{n_1}{D}\cdot
\frac{n_2}{D}\cdot
\cdots
\frac{n_m}{D} \,\,\,,
\eeqs
where the sum ${\begin{array}[t]{c}
{\displaystyle {\mathop{{\sum}'}}}\\
{\scriptstyle 1\leq {n_1} < {n_2} < \cdots < {n_m} \leq D-1}
\end{array}
}$
means that the variables $\{{n_i}\}$ run over integers which are not
divisible by any ${l_j}$,
\[
\begin{array}[t]{c}
{\displaystyle {\mathop{{\sum}'}}}\\
{\scriptstyle 1\leq {n_1} < {n_2} < \cdots < {n_m} \leq D-1}
\end{array}
=
\begin{array}[t]{c}
{\displaystyle {\mathop{{\sum}}}}\\
{\scriptstyle 1\leq {n_1} < {n_2} < \cdots < {n_m} \leq D-1}\\
{\scriptstyle {l_j} {\not{|}} {n_i}}
\end{array}\,\,\,.
\]
By introducing a new variable ${t:={\omega_1}(z)}$, the above
differential equation can be rewritten,
\beqs
&&{\del_{t}}\Phi (t)={C_t}\,\Phi  (t)\,\,\,,\,\,\,
{\Phi_{lm}}:={\int_{\gamma_m}}{\tilde{\alpha}_l}(t)\,\,\,,\\
&&{C_t}:=\left(
\begin{array}{ccccccc}
 0 & {\Kappa_0} &  &  &  &  & \mbox{\Large $O$} \\
   & 0 & {\Kappa_1}  &  &  &  &  \\
   &   & 0 & {\Kappa_2}  &  &  &  \\
   &   &   & \ddots  & \ddots &  &  \\
   &   &   &   & 0  & {\Kappa_{d-2}} &  \\
   &   &   &   &   & 0 & {\Kappa_{d-1}} \\
\mbox{\Large $O$} &   &   &   &   &   & 0
\end{array}
\right) \,\,\,,\\
&&{\tilde{\alpha}_0}(t):=\frac{1}{\varpi_0}{\alpha_0}\,\,\,,\\
&&{\tilde{\alpha}_l}(t):=
\frac{1}{{\Kappa}_{l-1}}{\del_{t}}
\frac{1}{{\Kappa}_{l-2}}{\del_{t}}\cdots {\del_{t}}
\frac{1}{{\Kappa}_{1}}{\del_{t}}
\frac{1}{{\Kappa}_{0}}{\del_{t}}
\left({\frac{\alpha_0}{\varpi_0}}\right)
\,\,\,,\,\,\,({1\leq l \leq d})\,\,\,,\\
&&{{\Kappa}_0}:={\del_{t}}
\,{\omega_1}=1\,\,\,,\,\,({t\equiv {\omega_1}})\\
&&{{\Kappa}_m}:=
{\del_{t}}\frac{1}{{\Kappa}_{m-1}}{\del_{t}}
\frac{1}{{\Kappa}_{m-2}}{\del_{t}}\cdots {\del_{t}}
\frac{1}{{\Kappa}_{1}}{\del_{t}}
\frac{1}{{\Kappa}_{0}}{\del_{t}}\,{\omega_{m+1}}
\,\,\,,\,\,\,({1\leq m \leq d-1})\,\,\,.
\eeqs
{}From this equation, we can read the action of the differential
operator ${\del_{t}}$
on the cohomology basis ${{\tilde{\alpha}_{l}}(t)}$,
\beqs
&&{\del_{t}}{\tilde{\alpha}_{j-1}}(t)={\Kappa_{j-1}}(t)\,
{\tilde{\alpha}_{j}}(t)\,\,\,,\,\,\,({1\leq j \leq d})\,\,\,,\\
&&{\del_{t}}{\tilde{\alpha}_{d}}(t)=0\,\,\,,\\
&&{\Kappa_{j-1}}(t)={\del_{t}}{\Phi_{j-1\,,\,j}}\,\,\,,\,\,\,({1\leq j \leq d})
\,\,\,.
\eeqs
When we define two point functions,
\[
\corr{{\tilde{\alpha}_{i}}{\tilde{\alpha}_{j}}}:=
{\int_{W}}{\tilde{\alpha}_{i}}\wedge {\tilde{\alpha}_{j}}\,\,\,,
\]
they satisfy relations,
\[
\corr{{\tilde{\alpha}_{i}}{\tilde{\alpha}_{j}}}:={\delta_{i+j,d}}
\corr{{\gamma^{\ast}_{i}}{\gamma^{\ast}_{j}}}\,\,\,,
\]
where the set ${\{{{\gamma^{\ast}_{m}}}\}}$ is the dual of the homology
d-cycles ${\{{\gamma_{m}}\}}$,
\[
{\int_{\gamma_{m}}}{\gamma^{\ast}_{n}}={\delta_{m,n}}\,\,\,.
\]
We translate these relations in the above B(W)-model into
the operator structures of the corresponding A(M)-model,
\beqs
&&{{\cal O}^{(1)}}{{\cal O}^{({j-1})}}=
{\Kappa_{j-1}}(t){{\cal O}^{(j)}}\,\,\,,\,\,\,({1\leq j \leq d})\,\,\,,\\
&&{{\cal O}^{(1)}}{{\cal O}^{({d})}}=0\,\,\,,\\
&&\corr{{{\cal O}^{(i)}}{{\cal O}^{(j)}}}=
{\delta_{i+j,d}}{\Eta_{ij}}\,\,\,,
\eeqs
for A(M)-model operators ${{{\cal O}^{(i)}}\in {H^{i,i}}(M)}\,,\,
({1\leq i \leq d})$.
%\newpage
The above operator product structure of the A(M)-model observables
is meaningful when one defines correlation functions in the following
way,
\beqs
&&\corr{{{\cal O}^{(1)}}{{\cal O}^{(j-1)}}\cdots }
:= \int {\cal D}[{X,\chi , \rho}]\,
{{\cal O}^{(1)}}{{\cal O}^{(j-1)}}\cdots {e^{-{L_A}}}\,\,\,,\\
&&\,\,\,\,\,\,{L_A}:=t{\int_{\Sigma}}{X^{\ast}}(e)+  {\int_{\Sigma}}{d^2}z \,
\ac{Q}{V}\,\,\,,
\eeqs
where $Q$ is a BRST charge of the A(M)-model and $V$ is given as,
\[
V:=it {g_{i\bar{\jmath}}}\,({{\rho^{\bar{\jmath}}_{z}}
{\del_{\bar{z}}}{X^{i}}+
{\rho^{j}_{\bar{z}}}
{\del_{z}}{X^{\bar{\imath}}}
})\,\,\,.
\]
Also the integral,
\[
{\int_{\Sigma}}{X^{\ast}}(e)=
{\int_{\Sigma}}{g_{i\bar{\jmath}}}\,
({ {\del_z}{X^i}{\del_{\bar{z}}}{X^{\bar{\jmath}}}-
{\del_{\bar{z}}}{X^i}{\del_z}{X^{\bar{\jmath}}} })\,
 dz \wedge d\bar{z} \,\,\,,
\]
is the pullback of the {\kae} form $e$ of M and equals to the degree
of the maps $X$. (This operator is a $2$-form version on the world
sheet of the local observable ${{\cal O}^{(1)}}$ ).
When we view this A(M)-model as a deformed theory from some
topological field theory, it is characteristic that we perturb the
original topological theory by adding only operators associated with
{\kae} forms of M. As for the (topological) selection rule for the
A(M)-model correlators, it depends on the degree of maps $X$ generally
because the virtual dimension (the ghost number anomaly) is given as,
\[
virdim= (dim\,M)\cdot (1-g)+{\int_{\Sigma}}{X^{\ast}}{c_1}(M)\,\,\,,
\]
where $g$ is the genus of the Riemann surface and ${c_1}(M)$ is the
first Chern class of M.
However for specific cases ${{c_1}=0}$ (Calabi-Yau cases), the virtual
dimension is independent of the degree of maps and then the selection
rule does not depend on the degree of maps.
Collecting these considerations, we can understand that the degree
conservation of A(M)-model operators in each fusion coincides with a
classical one for Calabi-Yau cases.
(Especially a relation ${{{\cal O}^{(1)}}{{\cal O}^{(d)}}=0}$
\,\,(d:=dim\,M) is satisfied). On the other hand, when we expand
fusion couplings $\{{{\Kappa}_{l}}\}$ of operators
${{{\cal O}^{(1)}}}$ and ${{{\cal O}^{(l)}}}$ with respect to an
indeterminate ${q:={e^{2\pi i t}}}$, they contain all non-negative
powers of $q$ generally because the selection rule of observables is
independent of the degree of maps.

Next let us consider a moduli space
of the Riemann surface (world sheet). The dimension of a moduli space
${{\cal M}_{g,s}}$ of a genus $g$ Riemann surface with $s$ punctures
is given as,
\[
dim\, {{\cal M}_{g,s}}=3(g-1)+s\,\,\,.
\]
Because we consider three point couplings ({${s=0}$}) on the sphere
({${g=0}$}), the degree of the world sheet moduli comes $3$ from the
positions of the operator insertions and ${-3}$ from the
{${SL}$}(2,{${\bc}$})-invariance of the ${{CP}^1}$ respectively.
Adding up these two contributions, we obtain the dimension of the moduli
space ${{\cal M}_{0,3}}$ of the Riemann surface,
\[
dim\,{{\cal M}_{0,3}}=3-3=0\,\,\,.
\]
Judging from this counting of the dimension only, we cannot understand
whether our systems (the Calabi-Yau matters) couple with the
topological gravity or not. In fact in our cases the systems do not
couple the gravity because we fix the positions of the operator
insertions and do not move them.
%%%%%%%%%%%%%%%%%%%%zzzzzzzzzzzzzzzz

Using this translation, we obtain three point functions ${{\Kappa_{l}}(t)}$
of the A(M)-model explicitly,
\beqs
&&{\Kappa_0}=1\,\,\,,\\
&&{\Kappa_{l}}=
{\del_{t}}\frac{1}{\Kappa_{l-1}}
{\del_{t}}\frac{1}{\Kappa_{l-2}}{\del_{t}}\cdots
{\del_{t}}\frac{1}{\Kappa_{2}}
{\del_{t}}\frac{1}{\Kappa_{1}}
{\del_{t}}\frac{1}{\Kappa_{0}}
{\del_{t}}\,{S_{l+1}}({{\tilde{x}_1}\,,\,{\tilde{x}_2}\,,\,\cdots \,,\,
{\tilde{x}_{l+1}}})\,\,\,,\,\,\,({1\leq l \leq d-1})\,\,\,,\\
&&{\tilde{x}_{n}}:=\frac{1}{n!}{{D}^{n}_{\rho}}\log {\tilde{\varpi}_0}
({z\,;\,\rho})
{\Biggr|_{\rho =0}}\,\,\,,\,\,\,
{{\dc}_{\rho}}:=\frac{1}{2\pi \,i }\cdot \frac{\del}{\del \rho}\,\,\,,\\
&&{\tilde{\vpi}_0} ({z,\rho}):= \sum^{\infty}_{m=0}
\frac{\Gamma ({D({m+\rho})+1})}{\Gamma ({D{\rho}+1})}\cdot
{\prod^{d+2}_{i=1}}
\frac{\Gamma ({{w_i}{\rho}+1})}{\Gamma ({{w_i}({m+\rho})+1})}
\,{z^{m+\rho}} \,\,\,,
\eeqs
where the function "${S_n}$" is the Schur function defined as the coefficients
in the following expansion,
\[
{\sum^{\infty}_{n=0}}{S_n}({x_1},{x_2},\cdots ,{x_n}) {u^n}:=
\exp \left({{\sum^{\infty}_{m=1}}{x_m}{u^m}}\right) \,\,\,.
\]
We write down expressions of these couplings ${\Kappa_{l}}$ in a series
with respect to a parameter ${{q:={e^{2\pi i\,t}}}\,\,,\,\,
t={S_1}({\tilde{x}_1})={\tilde{x}_1}}$,
\beqs
&&{\Kappa_l}=1+{\alpha_l}\,q+\mbox{\large $O$}({q^2})\,\,\,,\\
&&{\alpha}_l = \frac{D!}{\kw}\times
\,\Biggl[ {\tilde{A}}_{d+1-l}
-\left({{\displaystyle \sum^{D}_{l=1} \frac{D}{l}
-{\sum^{d+2}_{i=1}}{\sum^{w_i}_{{m_i}=1}} \frac{w_i}{m_i}}
}\right)
\, \Biggr] \,\,\,,\\
&&{{\tilde{A}}_m} :=
\begin{array}[t]{c}
{\displaystyle {\mathop {{\sum}'}}}\\
{\scriptstyle 1 \leq {m_1}<{m_2}< \cdots  <{m_n} \leq D-1 }
\end{array}
\frac{D-{m_1}}{m_1}\cdot \frac{D-{m_2}}{m_2}\cdot \cdots
\frac{D-{m_n}}{m_n} \,\,\,.
\eeqs
Then a d-point coupling of d operators {${{\cal O}^{(1)}}$}'s can be
calculated up to an overall normalization factor,
\beqs
&&\corr{{{\cal O}^{(1)}}{{\cal O}^{(1)}}\cdots {{\cal O}^{(1)}}}
={\Kappa_{1}}{\Kappa_{2}}\cdots {\Kappa_{d-2}}{\Eta_{1,d-1}}\\
&&=
1+\Biggl\{
\frac{D^D}{\pw}-2\cdot\frac{D!}{\kw}\\
&&\hspace{1.5cm}-d \cdot \frac{D!}{\kw}\times
{\Biggl[{{\sum^{D}_{l=1}} \frac{D}{l}-{\sum^{d+2}_{i=1}} {\sum^{w_i}_{m_i =1}}
\frac{w_i}{m_i}}\Biggr]}\Biggr\}
\cdot q+\mbox{\large $O$}({q^2}) \,\,\,.
\eeqs

\section{The Two-Parameter Family}

\pr
Secondly let us investigate the model (II).
The $G$-invariant parts of the Hodge structure of the mirror manifolds
$W=M/G$ are characterized by a set of homology d-cycles,
\beqs
{\{{{\gamma_0},{\gamma^{(1)}_1},{\gamma^{(2)}_1},\cdots \,
{\gamma^{(1)}_{d-1}},{\gamma^{(2)}_{d-1}},{\gamma_{d}}}\}}\,\,\,,
\eeqs
and a set of cohomology elements of $W$,
\beqs
&&{\{{{\alpha_0},{\alpha^{(1)}_1},{\alpha^{(2)}_1},\cdots ,
{\alpha^{(1)}_{d-1}},{\alpha^{(2)}_{d-1}},{\alpha_{d}}}\}}\,\,\,,\\
&&{\alpha^{({\bullet})}_{i}}\in {{\cal F}^{d-i}}=
{H^{d,0}}\oplus {H^{d-1,1}}\oplus \cdots \oplus {H^{d-i,i}}\,\,\,.
\eeqs
Especially we take the following elements as this basis,
\beqs
&&{\alpha_0}:=\Om ={\int_{\gamma}}\frac{d\mu}{\ps}\,\,\,\,\,
({\mbox{a holomorphic d form of $W$}})\,\,\,,\\
&&{\alpha^{(1)}_{l}}:={\Theta^{l}_{x}}\Om \,\,\,,
\,\,\,({1\leq l \leq d-1})\,\,\,,\\
&&{\alpha^{(2)}_{l}}:={\Theta^{l-1}_{x}}{\Theta_{y}}\Om \,\,\,,
\,\,\,({1\leq l \leq d-1})\,\,\,,\\
&&{\alpha_{d}}:=({{\Theta^{d}_{x}}-2\cdot {\Theta^{d-1}_{x}}{\Theta_{y}}})
\,\Om
\,\,\,,\\
&& x:=\frac{-2\phi}{{[{2(d+1)\,\psi}]}^{d+1}}\,\,,\,\,
y:=\frac{1}{{({2\phi})}^2}\,\,\,,\\
&&{\Theta_{x}}:=x\frac{\del}{\del {x}}\,\,,\,\,
{\Theta_{y}}:=y\frac{\del}{\del {y}}\,\,\,.
\eeqs
By using these elements, we can write a period matrix $P$
of the mirror $W$ in a
block form, which contains ${({d+1})\times ({d+1})}$ block matrices,
( ${4\,\,\,{1\times 1}}$ matrices, ${{2\cdot (d-1)}\,\,\,{1\times 2}}$
matrices, ${{2\cdot (d-1)}\,\,\,{2\times 1}}$
matrices, and
${{(d-1) \cdot (d-1)}\,\,\,2\times 2}$ matrices ).
The {$({l,m})$}-th block matrix ${P_{lm}}$ of $P$ is defined as,
%%%%%%%%%%444444444
\beqs
{P_{0m}}
&:=&
\left\{
\begin{array}{ll}
{\int_{\gamma_0}}{\alpha_0} & (m=0) \\
\left({{\int_{\gamma^{(1)}_{m}}}{\alpha_0}\,
{\int_{\gamma^{(2)}_{m}}}{\alpha_0}}\right) & (1\leq m \leq d-1) \\
{\int_{\gamma_d}}{\alpha_0} & (m=d)
\end{array}
\right.\,\,\,,\\
&=&
\left\{
\begin{array}{ll}
{\vpi_0} & (m=0) \\
({\vpi^{(1)}_{m}}\,{\vpi^{(2)}_{m}}) & (1\leq m \leq d-1) \\
{\vpi_d} & (m=d)
\end{array}
\right.\,\,\,,\\
{P_{lm}}&:=&
\left\{
\begin{array}{ll}
 \left(
  \begin{array}{l}
     {\int_{\gamma_{0}}}{\alpha^{(1)}_{l}} \\
      {\int_{\gamma_0}} {\alpha^{(2)}_{l}}
   \end{array}\right) & (m=0) \\
 \left(
  \begin{array}{ll}
     {\int_{\gamma^{(1)}_{m}}}{\alpha^{(1)}_{l}} &
{\int_{\gamma^{(2)}_{m}}}{\alpha^{(1)}_{l}} \\
     {\int_{\gamma^{(1)}_{m}}}{\alpha^{(2)}_{l}} &
{\int_{\gamma^{(2)}_{m}}}{\alpha^{(2)}_{l}}
   \end{array} \right) & (1\leq m \leq d-1) \\
  \left(
   \begin{array}{l}
     {\int_{\gamma_d}}{\alpha^{(1)}_{l}} \\
      {\int_{\gamma_d}}{\alpha^{(2)}_l}
   \end{array} \right) & (m=d)
\end{array}
\right.\,\,\,,\\
&=&
\left\{
\begin{array}{ll}
 \left(
  \begin{array}{l}
     {\th{x}}^{l} {\vpi_0} \\
      {\th{x}}^{l-1} {\th{y}}{\vpi_0}
   \end{array}\right) & (m=0) \\
 \left(
  \begin{array}{ll}
     {\th{x}}^{l} {\vpi^{(1)}_m} & {\th{x}}^{l} {\vpi^{(2)}_m} \\
      {\th{x}}^{l-1} {\th{y}}{\vpi^{(1)}_m} &
      {\th{x}}^{l-1} {\th{y}}{\vpi^{(2)}_m}
   \end{array} \right) & (1\leq m \leq d-1) \\
  \left(
   \begin{array}{l}
     {\th{x}}^{l} {\vpi_d} \\
      {\th{x}}^{l-1} {\th{y}}{\vpi_d}
   \end{array} \right) & (m=d)
\end{array}
\right.\,\,\,
\\
&&\hspace{4cm}(l=1,2,\cdots ,d-1)\,\,\,,\\
{P_{dm}}&:=&
\left\{
\begin{array}{ll}
 {\int_{\gamma_0}}{\alpha_d} & (m=0) \\
  \left(
   \begin{array}{ll}
  {\int_{\gamma^{(1)}_{m}}}{\alpha_d} &
  {\int_{\gamma^{(2)}_{m}}}{\alpha_d}
   \end{array}
  \right) & (1\leq m \leq d-1) \\
{\int_{\gamma_d}}{\alpha_d} & (m=d)
\end{array}
\right.\\
&=&
\left\{
\begin{array}{ll}
({{\th{x}}^d}-2{{\th{x}}^{d-1}}{\th{y}}) {\vpi_0} & (m=0) \\
  \left(
   \begin{array}{ll}
    ({{\th{x}}^d}-2{{\th{x}}^{d-1}}{\th{y}}) {\vpi^{(1)}_m} &
     ({{\th{x}}^d}-2{{\th{x}}^{d-1}}{\th{y}}) {\vpi^{(2)}_m}
   \end{array}
  \right) & (1\leq m \leq d-1) \\
({{\th{x}}^d}-2{{\th{x}}^{d-1}}{\th{y}}) {\vpi_d} & (m=d)
\end{array}
\right.\,\,\,.
\eeqs
We obtain a set of differential equations satisfied by
${\varpi^{({\bullet})}_{m}}\,\,\,\,({0\leq m \leq d})$,
\beq
&&{\cal D}_{{(1)}} {{\vpi}_m} (x,y)=0 \,\,\,,\,\,\,
{\cal D}_{{(2)}} {{\vpi}_m} (x,y)=0 \nonumber\,\,\,,\\
&&{\cal D}_{{(1)}} :={\th{x}}^{d-1} ({\th{x} -2 \th{y}}) \nonumber \\
&&\hspace{1.2cm} -{{(d+1)}^{d+1}}
x(\th{x} +\frac{d}{d+1})(\th{x} +\frac{d-1}{d+1})\cdots
(\th{x} +\frac{2}{d+1})(\th{x} +\frac{1}{d+1})
\label{eqn:od1}\,\,\,,\\
&&{\cal D}_{{(2)}} :={\th{y}}^2 -y({\th{x} -2\th{y}})
({\th{x} -2\th{y} -1}) \label{eqn:od2}\,\,\,.
\eeq
The $2d$ linear independent solutions are written down,
\beq
&&{\varpi_0}:={\hat{\varpi}_0}(x,y\,;\,{\rho_1},{\rho_2})
{\biggr|_{{\rho_1}={\rho_2}=0}}\,\,\,,\\
&&{{\vpi}^{(1)}_{l}}:=\frac{1}{l!}\,
{{\cal D}^{l}_{\rho_1}}{\hat{\vpi}_0}(x,y\,;\,{\rho_1},{\rho_2})
{\mbox{\Large $|$}_{{\rho_1}={\rho_2}=0}}\,\,\,,\,\,
(l=1,2,\cdots ,d-1)\,\,\,,\label{eqn:kai1}\\
&&{{\vpi}^{(2)}_{l}}:=\frac{1}{{(l-1)}!}
{{\cal D}^{l-1}_{\rho_1}}
{{\cal D}_{\rho_2}}{\hat{\vpi}_0}(x,y\,;\,{\rho_1},{\rho_2})
{\mbox{\Large $|$}_{{\rho_1}={\rho_2}=0}}\,\,\,,\,\,
(l=1,2,\cdots ,d-1)\,\,\,,\label{eqn:kai2}\\
&&{{\vpi}_{d}}:=
\frac{1}{d!}
\left({2\,{{\cal D}^{d}_{\rho_1}}+
d\cdot {{\cal D}^{d-1}_{\rho_1}}
{{\cal D}_{\rho_2}}}\right)
{\hat{\vpi}_0}(x,y\,;\,{\rho_1},{\rho_2})
{\mbox{\Large $|$}_{{\rho_1}={\rho_2}=0}}\,\,\,,\label{eqn:kai3}
\eeq
where
\beqs
&&{\hat{\vpi}_0} ({x,y\,,\,{\rho_1},{\rho_2}})
:={\sum_{m,n\geq 0}}
\frac{\Gamma ({(d+1)({m+{\rho_1}})+1})}{\Gamma ({(d+1){\rho_1}+1})}\\
&& \hspace{1cm} \times
{{\left[{\frac{\Gamma ({1+{\rho_1}})}
{\Gamma ({m+1+{\rho_1}})}}\right]}^{d}}
\times
{{\left[{\frac{\Gamma ({1+{\rho_2}})}
{\Gamma ({n+1+{\rho_2}})}}\right]}^{2}} \\
&& \hspace{1cm} \times
{\left[{\frac{\Gamma ({{\rho_1}-2{\rho_2}+1})}
{\Gamma ({m-2n+{\rho_1}-2{\rho_2}+1})}}\right]}
{x^{m+{\rho_1}}}{y^{n+{\rho_2}}} \,\,\,.
\eeqs
\beqs
{{\cal D}_{\rho_i}}:&=&\frac{1}{2\pi i}\,\frac{\del}{\del {\rho_i}}\,\,\,.
\eeqs
The property of the set of homology cycles is characterized by
monodromy transformations around the points ${x=0}$ or ${y=0}$.
When we turn around the points ${x=0}$ or ${y=0}$,
\[
x\rightarrow {e^{2\pi i}}x \,\,\,\mbox{or}\,\,\,
y\rightarrow {e^{2\pi i}}y\,\,\,,
\]
the period matrix $P$ of $W$ is transformed into a form,
\[
{P_{lm}}\rightarrow {\sum_{n}}{P_{ln}}{T^{(i)}_{nm}}
\,\,\,,\,\,\,\left\{
          \begin{array}{lcl}
             i=1 & & ({x=0}) \\
             i=2 & & ({y=0})
           \end{array}\right.\,\,\,,
\]
where the ${2d\times 2d}$ matrices ${T^{(1)}}$ and ${T^{(2)}}$ are
expressed as,
\beqs
\left\{
\begin{array}{llll}
{T^{(1)}} & = & \exp ({N^{(1)}}) & (\mbox{around $x=0$}) \\
{T^{(2)}} & = & \exp ({N^{(2)}}) & (\mbox{around $y=0$})
\end{array}
\right.\,\,\,,
\eeqs
\begin{eqnarray}
{N^{(1)}}:=
\begin{array}{r@{}l}
& \begin{array}{ccccccc}
{\scriptscriptstyle {\overbrace{}^{1}} }\!\!\! &
{\scriptscriptstyle {\overbrace{}^{2}} }\!\!\! &
{\scriptscriptstyle {\overbrace{}^{2}} }\!\!\! &
{\scriptscriptstyle {\overbrace{}^{2}} }\!\!\! &
\cdots &
{\scriptscriptstyle {\overbrace{}^{2}} }\!\!\! &
{\scriptscriptstyle {\overbrace{}^{1}} }
\end{array} \\
\begin{array}{l}
1\{ \\
2\{ \\
2\{ \\
\vdots \\
2\{ \\
2\{ \\
1\{ \\
\end{array} & \left(
\begin{array}{ccccccc}
0 & {e^{(1)}_{1}} &   &  &   & {\mbox{\Large $O$}} & \\
  & 0             & I &  &   &  & \\
  &  & 0 & I &  & & \\
  &  &   & \ddots & \ddots & & \\
  &  &   &   & 0 & I & \\
  &  &   &   &    &  0 &  {{e}^{(2)}_{1}}\\
  & {\mbox{\Large $O$}} &   &   &        & & 0
\end{array} \right)
\end{array}\,\,\,,
\end{eqnarray}
%%%%%% bmatrix
\begin{eqnarray}
{N^{(2)}}:=
\begin{array}{r@{}l}
& \begin{array}{ccccccc}
{\scriptscriptstyle {\overbrace{}^{1}} }\!\!\! &
{\scriptscriptstyle {\overbrace{}^{2}} }\!\!\! &
{\scriptscriptstyle {\overbrace{}^{2}} }\!\!\! &
{\scriptscriptstyle {\overbrace{}^{2}} }\!\!\! &
\cdots &
{\scriptscriptstyle {\overbrace{}^{2}} }\!\!\! &
{\scriptscriptstyle {\overbrace{}^{1}} }
\end{array} \\
\begin{array}{l}
1\{ \\
2\{ \\
2\{ \\
\vdots \\
2\{ \\
2\{ \\
1\{ \\
\end{array} & \left(
\begin{array}{ccccccc}
0 & {e^{(1)}_{2}} &   &  &   & {\mbox{\Large $O$}} & \\
  & 0             & I' &  &   &  & \\
  &  & 0 & I' &  & & \\
  &  &   & \ddots & \ddots & & \\
  &  &   &   & 0 & I' & \\
  &  &   &   &    &  0 &  {{e}^{(2)}_{2}}\\
  & {\mbox{\Large $O$}} &   &   &        & & 0
\end{array} \right)
\end{array}\,\,\,,
\end{eqnarray}
where the matrices in the blocks are given as,
\beqs
\begin{array}{cclccclc}
{e^{(1)}_{1}} & = &
   \left( \begin{array}{cc}
            1 & 0
          \end{array} \right) & , & {e^{(2)}_{1}} & = &
     \left( \begin{array}{c}
                2 \\
                1
            \end{array} \right) & , \\
{e^{(1)}_{2}} & = &
   \left( \begin{array}{cc}
            0 & 1
          \end{array} \right) & , & {e^{(2)}_{2}} & = &
       \left( \begin{array}{c}
                  1 \\
                  0
               \end{array} \right) & , \\
I & = &
   \left( \begin{array}{cc}
              1 & 0 \\
              0 & 1
          \end{array} \right) & , & I' & = &
   \left( \begin{array}{cc}
               0 & 1 \\
               0 & 0
           \end{array} \right) & .
\end{array}
\eeqs
A simple calculation shows several relations,
\beqs
&& {\left\{{N^{(1)}}\right\}}^{d+1}
=0\,\,\,,\,\,\,
{\left\{{N^{(2)}}\right\}}^{2} =0
\,\,\,,\,\,\,
{\left\{{N^{(1)}}\right\}}^{d}
-2\cdot {\left\{{N^{(1)}}\right\}}^{d-1} \cdot
{\left\{{N^{(2)}}\right\}} =0\,\,\,,\\
&& \cm{N^{(1)}}{N^{(2)}} =0\,\,\,,
\eeqs
and we obtain an evidence of the maximally unipotent
monodromy (the nilpotent properties ${{({N^{(1)}})}^{d+1}}=0 $,
${{({N^{(2)}})}^{2}}=0 $ show that maximally unipotent monodromy is realized at
the point ${({x\,,\,y})=({0\,,\,0})}$) and the homology cycles seem to be
chosen appropriately.
Let us introduce mirror maps in the model (II),
\beq
t({x\,,\,y}):=\frac{\varpi^{(1)}_{1}}{\varpi_0}\,\,\,,\,\,\,
s({x\,,\,y}):=\frac{\varpi^{(2)}_{1}}{\varpi_0}\,\,\,.\label{eqn:mirror}
\eeq
These maps behave under the transformations around ${x=0}$ or ${y=0}$ as,
\beqs
&& t({e^{2\pi \,i}}x,y)=t(x,y)+1\,\,\,,\,\,\,
t(x,{e^{2\pi \,i}}y)=t(x,y)\,\,\,,\\
&& s(x,{e^{2\pi \,i}}y)=s(x,y)+1\,\,\,,\,\,\,
s({e^{2\pi \,i}}x,y)=s(x,y)\,\,\,.
\eeqs
Because of these properties,
one may regard these maps as the coordinates of the {\kae} moduli
space, which are defined modulo some integer shifts in the physical
situations. Also the elements in the
zero-th row of the period matrix ${\Phi}$ are given by the
following functions,
\beqs
&&{\omega_0}=1\,\,\,,\\
&&{\omega^{(1)}_{l}} \equiv
\frac{\varpi^{(1)}_{l}}{\varpi_0}=
{S_l}
({{\tilde{x}_1}\,,\,{\tilde{x}_2}\,,\,\cdots \,,\,{\tilde{x}_l}})\\
&&={\sum^{l}_{m=0}} {a_{l-m}}\frac{t^m}{m!}
\,\,\,,\,\,\,(l=1,2,\cdots ,d-1)\,\,\,,\\
&&{\omega^{(2)}_{l}} \equiv
\frac{\varpi^{(2)}_{l}}{\varpi_0}=
s\cdot {S_{l-1}}
({{\tilde{x}_1}+{\tilde{y}_1}\,,\,
{\tilde{x}_2}+{\tilde{y}_2}\,,\,\cdots \,,\,
{\tilde{x}_{l-1}}+{\tilde{y}_{l-1}}})\\
&&={\sum^{l}_{m=1}}
{a_{l-m}}\frac{{t^{m-1}}\cdot s}{{(m-1)}!}+{\sum^{l}_{m=2}}
{c_{l-m}}\frac{t^{m-2}}{{(m-2)}!}
\,\,\,,\,\,\,(l=1,2,\cdots ,d-1)\,\,\,,\\
&&{\omega_{d}} \equiv
\frac{\varpi_{d}}{\varpi_0}=
2\cdot {S_d}
({{\tilde{x}_1}\,,\,{\tilde{x}_2}\,,\,\cdots \,,\,{\tilde{x}_d}})+
s\cdot {S_{d-1}}
({{\tilde{x}_1}+{\tilde{y}_1}\,,\,
{\tilde{x}_2}+{\tilde{y}_2}\,,\,\cdots \,,\,
{\tilde{x}_{d-1}}+{\tilde{y}_{d-1}}})\,\,\,,\\
&& \left\{
\begin{array}{llll}
{\tilde{x}_m} & := & \frac{1}{m!}
{{\dc}^{m}_{\rho_1}}{\log}{\hat{\varpi}_0}(x,y\,;\,  {\rho_1},{\rho_2})
{\biggr|_{{\rho_1}={\rho_2}=0}} & , \\
{\tilde{y}_m} & := & \frac{1}{m!}
{{\dc}^{m}_{\rho_1}}{\log}
\left[
{{\dc}_{\rho_2}}{\hat{\varpi}_0}(x,y\,;\,{\rho_1},{\rho_2})
\right] {\biggr|_{{\rho_1}={\rho_2}=0}} & , \\
{\hat{y}_m} & := & \frac{1}{m!}
{{\dc}^{m}_{\rho_1}}{{\dc}_{\rho_2}}{\log}
{\hat{\varpi}_0}(x,y\,;\,  {\rho_1},{\rho_2})
{\biggr|_{{\rho_1}={\rho_2}=0}} & ,
\end{array}
\right.\\
&&{a_k}:={\sum_{n\geq 1}}{\sum_{{\{\sharp\,\,1}\}}}
\frac{1}{n!}{\tilde{x}_{m_1}}{\tilde{x}_{m_2}}\cdots {\tilde{x}_{m_n}}\,\,\,,
\,\,\,
{c_k}:={\sum_{n\geq 0}}{\sum_{\{{\sharp\,\,2}\}}}
\frac{1}{n!}{\tilde{x}_{m_1}}{\tilde{x}_{m_2}}\cdots {\tilde{x}_{m_n}}
{\hat{y}_{m_{n+1}}}
\,\,\,,\\
&& \{{\sharp\,\,1}\} :=\{{{m_1}\geq 2\,,\,{m_2}\geq 2\,,\,\cdots \,,\,
{m_n}\geq 2\,,\,{m_1}+{m_2}+\cdots +{m_n}=k}\}\,\,\,,\\
&& \{{\sharp\,\,2}\} :=\{{{m_1}\geq 2\,,\,{m_2}\geq 2\,,\,\cdots \,,\,
{m_n}\geq 2\,,\,{m_{n+1}}\geq 1\,,\,
{m_1}+{m_2}+\cdots +{m_n}+{m_{n+1}}=k+1}\}\,\,\,.
\eeqs
Next multiplying some block lower triangular matrix to the period matrix $P$
from the left, we obtain a block upper triangular period
matrix $\Phi $ which has unit matrices in its block diagonal parts,
\beqs
\Phi =\left(
\begin{array}{ccccccccc}
1 & \om{1}^{({\bullet})} & \om{2}^{({\bullet})} & \om{3}^{({\bullet})} &
\om{4}^{({\bullet})} & \cdots & \om{d-2}^{({\bullet})} &
\om{d-1}^{({\bullet})} & \om{d} \\
  & I & \hm{1}{2} & \hm{1}{3} & \hm{1}{4} & \cdots &
 \hm{1}{d-2} & \hm{1}{d-1} & u^{(1)} \\
  &   & I & \hm{2}{3} & \hm{2}{4} & \cdots &
 \hm{2}{d-2} & \hm{2}{d-1} & u^{(2)} \\
  &   &  & I & \hm{3}{4} & \cdots &
 \hm{3}{d-2} & \hm{3}{d-1} & u^{(3)} \\
  &   &  &   & \ddots & \ddots &
 \vdots & \vdots & \vdots \\
  &   &  &   &        & I &
\hm{d-3}{d-2} & \hm{d-3}{d-1} & u^{(d-3)} \\
  &   &  &   &        &   &
I  & \hm{d-2}{d-1} & u^{(d-2)} \\
  &   &  &   &        &   &
  & I  & u^{(d-1)} \\
{\mbox{\Large $O$}}  &   &  &   &        &   &
  &   & 1
\end{array}
\right)\,\,\,.
\eeqs
%%%%%%%%%%%%%%%%%%%%%%%%%%%%%%%%%%%%%%%OA%%%%%%%%kkkkkkkkkkkkkkk
%%%%%%%%%%%%%%%%%%%%%%%%%%%%%%%%%%%%%%%%%%%%%%%
%%%%%%%%%%%%%%%%%%%%%%%%%%%%%%%%%%%%%%%%%%%%%%%
%%\newpage
%%%%%%%%%%%%%
%%%%%%%%%%%%%%%sssssssssssssssssssssss%%%%%%%%%%%%%%%%%%%%%%%%
The components in the first row are defined
by using the ${{\om{l}}^{(i)}}$,
\[
{\om{l}}^{({\bullet})} :=\left(
\begin{array}{ll}
{\om{l}}^{(1)} & {\om{l}}^{(2)}
\end{array}
\right) \,\,\,,\,\,\,(l=1,2,\cdots ,d-1)\,\,\,.
\]
The blocks ${\hm{m}{l}}$ can be written as,
\beqs
&&\hm{m}{l} ={U_{l-m}}+{\sum^{l-m-1}_{n=0}}{A^{(m)}_{n}}
{U_{l-m-n-1}}\,\,\,,\\
&&\hspace{2cm} (l>m\,;\,m=1,2,\cdots ,d-2\,;\,2\leq l \leq d-1 )\,\,\,,\\
&&{U_n}:=\left(
\begin{array}{cc}
\frac{t^n}{n!} & \frac{{t^{n-1}}\cdot s}{{(n-1)}!} \\
0 & \frac{t^n}{n!}
\end{array}
\right)\,\,\,,\,\,\,(n=1,2,\cdots )\,\,\,,\\
&&{U_0}:=\left(
\begin{array}{cc}
1 & 0 \\
0 & 1
\end{array}
\right) =I \,\,\,,
\eeqs
where the matrices ${A^{(m)}_{n}}$ are defined iteratively,
\beqs
&& {A^{(1)}_{n}}:=\left(
\begin{array}{cc}
{a_{n+1}}+{\del_t}{a_{n+2}} & {c_{n-1}}+{\del_t}{c_n} \\
{\del_s}{a_{n+2}} & {a_{n+1}}+{\del_s}{c_n}
\end{array}
\right)\,\,\,,\,\,\,(n=1,2,\cdots)\,\,\,,\\
&& {A^{(1)}_{0}}:=\left(
\begin{array}{cc}
{\del_t}{a_{2}} & {\del_t}{c_0} \\
{\del_s}{a_{2}} & {\del_s}{c_0}
\end{array}
\right)\,\,\,,\\
&&{A^{(m+1)}_{n}}:=
{({I+{\del_t}{A^{(m)}_{0}}})}^{-1}
\cdot {({{A^{(m)}_{n}}+{\del_t}{A^{(m)}_{n+1}}})}\,\,\,,\,\,\,
(m\geq 1)\,\,\,.
\eeqs
Also the matrices ${u^{(l)}}$ can be written as,
\beqs
&&{u^{(1)}}=\left(
\begin{array}{l}
{\del_t}{\omega_d} \\
{\del_s}{\omega_d}
\end{array}
\right)\,\,\,,\\
&&{u^{(l)}}:=
{\left({{\del_t}{\hm{l-1}{l}}}\right)}^{-1}
{\del_t}{u^{(l-1)}} \,\,\,,\,\,\,(l=2,3,\cdots ,d-1)\,\,\,.
\eeqs
%%%%%%%%%%%%%%%%%%%aaaaaaaaaaaaaaaaaaaaaa
Let us investigate differential equations for the resulting
period $\Phi$.
Note that the differential
equations of the original period $P$,
\beq
&& \left({\th{x} -{A_1} }\right) \,{P}=0\,\,\,,\nonumber \\
&& \left({\th{y} -{A_2} }\right) \,{P}=0\,\,\,,\label{eqn:set1}
\eeq
where both matrices ${A_1},{A_2}$ have non-vanishing elements
only at the lower triangular blocks, the diagonal blocks
and the first upper triangular blocks from the
diagonal blocks,
\beq
{A_i}:=\left(
\begin{array}{ccccccc}
\ast & \ast &  &  &  &  & \mbox{\Large $O$} \\
\ast & \ast & \ast  &  &  &  &  \\
\ast & \ast & \ast  &  \ast &  &  &  \\
\vdots & \vdots & \vdots  & \ddots  & \ddots &  &  \\
\ast & \ast & \ast  &  \cdots & \ast & \ast &  \\
\ast & \ast & \ast  &  \cdots & \ast & \ast & \ast \\
\ast & \ast & \ast  &  \cdots & \ast & \ast & \ast
\end{array}
\right) \,\,\,.\label{eqn:block1}
\eeq
This can be easily understood from the forms of the original
differential equations.
When we carry out the sweeping-out method on the equations (\ref{eqn:set1}),
the period matrix $P$ is changed into the $\Phi$ by multiplying
some lower triangular matrix $g$,
\beq
&&\Phi =g\,P \,\,\,,\nonumber \\
&&g=\left(
\begin{array}{ccccccc}
\ast & & & & & & \mbox{\Large $O$} \\
\ast & \ast & & & & & \\
\ast & \ast & \ast & & & & \\
\vdots & \vdots & \vdots & \ddots & & & \\
\ast & \ast & \ast & \cdots & \ast & & \\
\ast & \ast & \ast & \cdots & \ast & \ast & \\
\ast & \ast & \ast & \cdots & \ast & \ast & \ast
\end{array}
\right) \,\,\,.\label{eqn:block2}
\eeq
Then equations (\ref{eqn:set1}) are rewritten as,
\beq
\left[{\th{i} -\left({g\,{A_i}\,{g^{-1}}-g\,{\th{i}}\,{g^{-1}}}\right)
}\right] \,\Phi =0\,\,\,,\,\,\,(i=1,2\,;\,{\th{1}}:={\th{x}}\,,\,
{\th{2}}:={\th{y}})\,\,\,.
\eeq
That is to say, the matrices ${A_i}$
 are transformed as components of
some gauge connection $A$,
\[
A:={A_1}d\,{\log x}+{A_2}d\,{\log y}\,\,\,,
\]
and the transformation by $g$ can be regarded as a sort of gauge
transformation.
{}From the forms of ${A_i}$ and $g$ (\ref{eqn:block1},\ref{eqn:block2}),
we find that the form of ${A^{g}_{i}}:=g\,{A_i}\,{g^{-1}}-g\,\th{i}\,{g^{-1}}$
 have the similar form as the ${A_i}$,
\beq
{A^g_i}:=\left(
\begin{array}{ccccccc}
\ast & \ast &  &  &  &  & \mbox{\Large $O$} \\
\ast & \ast & \ast  &  &  &  &  \\
\ast & \ast & \ast  &  \ast &  &  &  \\
\vdots & \vdots & \vdots  & \ddots  & \ddots &  &  \\
\ast & \ast & \ast  &  \cdots & \ast & \ast &  \\
\ast & \ast & \ast  &  \cdots & \ast & \ast & \ast \\
\ast & \ast & \ast  &  \cdots & \ast & \ast & \ast
\end{array}
\right) \,\,\,.\label{eqn:block3}
\eeq
On the other hand, the resulting period matrix $\Phi$ has the upper
triangular form and the matrix ${\th{i} \Phi}$ is also some
upper triangular matrix,
\beqs
\th{i} \Phi =\left(
\begin{array}{ccccccc}
 0   & \ast & \ast &  \ast  & \cdots & \ast & \ast \\
     &  0   & \ast &  \ast  & \cdots & \ast & \ast \\
     &      &  0   &  \ast  & \cdots & \ast & \ast \\
     &      &      & \ddots & \ddots & \vdots & \vdots \\
     &      &      &        &  0     & \ast & \ast \\
     &      &      &        &        &  0   & \ast \\
\mbox{\Large $O$} &      &      &        &      &     &  0
\end{array}
\right)\,\,\,.
\eeqs
Comparing the both hand sides of the equations,
\[
\th{i} \Phi ={A^{g}_{i}}\Phi \,\,\,,
\]
we find that the ${A^{g}_{i}}$ must have the following form,
\beq
{A^{g}_{i}}=\left(
\begin{array}{ccccccc}
 0 & \ast &  &  &  &  & \mbox{\Large $O$} \\
   & 0 & \ast  &  &  &  &  \\
   &   & 0 & \ast  &  &  &  \\
   &   &   & \ddots  & \ddots &  &  \\
   &   &   &   & 0  & \ast &  \\
   &   &   &   &   & 0 & \ast \\
\mbox{\Large $O$} &   &   &   &   &   & 0
\end{array}
\right) \,\,\,.\label{eqn:block4}
\eeq
Lastly we change the variables $(x,y)$ into the variables $(t,s)$ defined in
({\ref{eqn:mirror}}),
we obtain the results,
\beq
&&{\del_{t_i}}\Phi ={\Kappa_i}\Phi \,\,\,,\,\,\,(i=1,2\,;\,{t_1}:=t\,,\,
{t_2}:=s)\,\,\,\nonumber \\
&&{\Kappa_i}={\del_{t_i}}{\sf F}\,\,\,,\nonumber \\
&&{\sf F}:=\left(
\begin{array}{ccccccc}
 0 & {\om{1}}^{({\bullet})} &  &  &  &  & \mbox{\Large $O$} \\
   & 0 & \hm{1}{2}  &  &  &  &  \\
   &   & 0 & \hm{2}{3}  &  &  &  \\
   &   &   & \ddots  & \ddots &  &  \\
   &   &   &   & 0  & \hm{d-2}{d-1} &  \\
   &   &   &   &   & 0 & {u^{(d-1)}} \\
\mbox{\Large $O$} &   &   &   &   &   & 0
\end{array}
\right) \,\,\,.\label{eqn:block5}
\eeq
That is to say, three point couplings ${\Kappa^{(l)}_{i}}$
defined by fusions,
\[
{{\cal O}^{(1)}_{i}}\cdot {{\cal O}^{(l)}_{j}}=
{\left({\Kappa^{(l)}_{i}}\right)}^{k}_{j}
{{\cal O}^{(l+1)}_{k}} \,\,\,,
\]
are obtained,
\beqs
{{\left({\Kappa^{(l)}_{i}}\right)}^{k}_{j}}=
{{\left({{\del_{t_i}}\hm{l}{l+1} }\right)}_{jk} }\,\,\,.
\eeqs
%%%%%%%%%%%bbbbbbbbbbbbbbb
Because $(l,l+1)$-th block can be represented as,
\[
\hm{l}{l+1} =\left(
\begin{array}{cc}
t & s \\
0 & t
\end{array}
\right) +{A^{(l)}_{0}} \,\,\,,\,\,\,(l=1,2,\cdots ,d-2)\,\,\,,
\]
we get three point functions in the matrix forms,
\beqs
&& {\Kappa}^{(l)}_{t} :=
{\del_t}{\hm{l}{l+1}}=\left(
\begin{array}{cc}
1 & 0 \\
0 & 1
\end{array}
\right) +{\del_t}{A^{(l)}_{0}}\,\,\,,\\
&& {\Kappa}^{(l)}_{s} :=
{\del_s}{\hm{l}{l+1}}=\left(
\begin{array}{cc}
0 & 1 \\
0 & 0
\end{array}
\right) +{\del_s}{A^{(l)}_{0}}\,\,\,.
\eeqs
%\newpage
These couplings are represented graphically,
\beqs
&&{\Kappa}^{(l)}_{t} =
\mbox{
\setlength{\unitlength}{1mm}
\begin{picture}(90,23)(0,22.5)
\put(25,15){\line(1,0){40}}
\put(45,15){\line(0,1){15}}
\put(14,12){\makebox(6,6){${{\cal O}^{(l)}}$}}
\put(70,12){\makebox(6,6){${{\cal O}^{(l+1)}}$}}
\put(42,33){\makebox(6,6){${{\cal O}^{(1)}_{t}}$}}
\put(25,15){\circle*{2}}
\put(65,15){\circle*{2}}
\put(45,30){\circle*{2}}
\end{picture}
}\,\,\,,\\
&&{\Kappa}^{(l)}_{s} =
\mbox{
\setlength{\unitlength}{1mm}
\begin{picture}(90,23)(0,22.5)
\put(25,15){\line(1,0){40}}
\put(45,15){\line(0,1){15}}
\put(14,12){\makebox(6,6){${{\cal O}^{(l)}}$}}
\put(70,12){\makebox(6,6){${{\cal O}^{(l+1)}}$}}
\put(42,33){\makebox(6,6){${{\cal O}^{(1)}_{s}}$}}
\put(25,15){\circle*{2}}
\put(65,15){\circle*{2}}
\put(45,30){\circle*{2}}
\end{picture}
}\,\,\,,\\
\eeqs
where the symbols ${{\cal O}^{(1)}_{t}}$, ${{\cal O}^{(1)}_{s}}$
 stand for the injection of some charge $(1,1)$ operators
associated with the {\kae} forms ${J_t},{J_s}$ coupled with $t, s$
respectively.
In the above figure, charge $(l,l)$ operators ${{\cal O}^{(l)}}$ and the charge
$(1,1)$ operators ${{\cal O}^{(1)}_{t,s}}$ are injected and fuse together.
Then some resulting operators ${{\cal O}^{(l+1)}}$ with charge $(l+1,l+1)$
are constructed. In this case, there are just two operators with each
definite charge $(l,l)\,\,(l=1,2,\cdots ,d-1)$.
Thus we can represent the couplings which have one of the fixed charge $(1,1)$
operators ${{\cal O}^{(1)}_{t}}$, ${{\cal O}^{(1)}_{s}}$
as 2 by 2 matrices ${({\Kappa^{(l)}_t})}$,
${({\Kappa^{(l)}_s})}$\,\,$(l=1,2,\cdots ,d-1)$.
After straightforward calculations, we obtain series expansions of these
couplings explicitly,
\beqs
&&{\Kappa^{(l)}_{t}}=\left(
\begin{array}{cc}
1 & 0 \\
0 & 1
\end{array}
\right) +{q_1}
\left(
\begin{array}{cc}
{\tilde{a}^{(l+1)}_{1}} & {\tilde{a}^{(l)}_{2}} \\
0  & {\tilde{a}^{(l)}_{1}}
\end{array}
\right) +{\mbox{\large $O$}}({q^2})\,\,\,,\\
&&{\Kappa^{(l)}_{s}}=\left(
\begin{array}{cc}
0 & 1 \\
0 & 0
\end{array}
\right) +{q_2}
\left(
\begin{array}{cc}
0  & 0 \\
1  & -2
\end{array}
\right) +{\mbox{\large $O$}}({q^2})\,\,\,,\\
&&{\tilde{a}^{(l)}_1} :={{(d+1)}!}\cdot
\Biggl[
-\left({\displaystyle  {\sum^{d+1}_{n=2}} \frac{d+1}{n} }\right) \\
&&\hspace{2cm}
+\begin{array}[t]{c}
{\displaystyle {\mathop {{\sum}}}}\\
{\scriptstyle 1 \leq {m_1}<{m_2}< \cdots  <{m_l} \leq d}
\end{array}
\frac{d+1-{m_1}}{m_1}\cdot \frac{d+1-{m_2}}{m_2}\cdot \cdots
\frac{d+1-{m_l}}{m_l}
\Biggr] \,\,\,,\\
&&{\tilde{a}^{(l)}_2} :={{(d+1)}!}\cdot 2\,
\Biggl[
-1
+\begin{array}[t]{c}
{\displaystyle {\mathop {{\sum}}}}\\
{\scriptstyle 1 \leq {m_1}<{m_2}< \cdots  <{m_l} \leq d}
\end{array}
\frac{d+1-{m_1}}{m_1}\cdot \frac{d+1-{m_2}}{m_2}\cdot \cdots
\frac{d+1-{m_l}}{m_l}
\Biggr] \,\,\,,\\
&&\hspace{5cm}(l=1,2,\cdots ,d-1)\,\,\,,\\
&& {q_1} :=\exp (2\pi \,i\, t)\,\,\,,\,\,\,
{q_2 }:=\exp (2\pi \,i\, s)\,\,\,\,\,.
\eeqs
The coefficients in the series expansions of these three point functions,
\beqs
{{({\Kappa^{(l-1)}_a})}_{b\, c}}&:=&
\corr{{{\cal O}^{(1)}_{a}}(0)\,{{\cal O}^{(l-1)}_{b}}(1)\,
{{\cal O}^{(d-l)}_{c}}({\infty})}\\
&=& \corr{{\phi}[{e^{(a)}_{1}}](0)\,{\phi}[{e^{(b)}_{l-1}}](1)\,
{\phi}[{e^{(c)}_{d-l}}]({\infty})
}\,\,\,,
\eeqs
 correspond to the number of holomorphic maps ${X^{i}}$
with the conditions,
\beqs
&&{{\cal O}^{(m)}_a}(P) \equiv {\phi}[{e^{(a)}_{m}}](P)\,\,\,,\,\,\,
{e^{(a)}_{m}}\in {H^{m,m}_d}\,\,,\,\,P\in \Sigma\,\,\,,\\
&& X(0)\in P.D.({e^{(a)}_{1}})\,\,\,,\\
&& X(1)\in P.D.({e^{(b)}_{l-1}})\,\,\,,\\
&& X({\infty})\in P.D.({e^{(c)}_{d-l}})\,\,\,,
\eeqs
where the $\phi $ maps cohomology elements ${e_m}$ of $M$ to some
A(M)-model operators ${{\cal O}^{(m)}}$.
%%%%%%%%%%%%%ccccccccccccccccc
The previous matrix ${\sf F}$ (\ref{eqn:block5}) can be thought of as a sort
of a generating function of the three point functions ${\Kappa_i}$.
This result is quite fascinating. All we have done seem to be a sort
of the Miura transformation in two-parameter case. From that viewpoint,
the resulting period matrix $\Phi$ is the positive root part of the matrix
$P$ in the usual Gauss decomposition, and the three point couplings are
associated to the Cartan parts of the Gauss decomposition.
We make a remark.
For the Calabi-Yau three-folds, the three point functions have charge one
fields only, and they are symmetric with respect to these three indices.
For that reason, one can integrate this function ${\sf F}$ more twice only
for the three-folds.

Next let us consider integrable conditions.
Note that we have all explicit forms of solutions of the equations and know
all components of the period matrix exactly in series formula.
Obviously integrable conditions should exist,
\beq
\cm{{\del_{t_i}}-{\Kappa_i}}{{\del_{t_j}}-{\Kappa_j}}=0 \,\,\,.
\label{eqn:kaseki}
\eeq
{}From this relation (\ref{eqn:kaseki}) and the equation
${\Kappa_i}={\del_i}{\sf F}$, we have some sort of
associativities among these couplings,
\beq
\cm{\Kappa_i}{\Kappa_j}=0 \,\,\,.\label{eqn:asso}
\eeq
They can be rewritten in components of the matrices
${{\Kappa_i}\,\,({i=1,2})}$ as,
\[
{\Kappa^{(l)}_{t_i}}{\Kappa^{(l+1)}_{t_j}}-
{\Kappa^{(l)}_{t_j}}{\Kappa^{(l+1)}_{t_i}}
=0\,\,\,.
\]
This relation tells us that any correlation functions are
independent of the positions of the insertion of the charge one
operators ${{\cal O}^{(1)}_{j}}$ and this property is reasonable from the
physical point of view.
%%%%%%%%%%%%%%eeeeeeeeeeeeeeeeeeeeee
Now we consider d-point functions of charge $(1,1)$ operators
${{\cal O}^{(1)}_{t}}$, ${{\cal O}^{(1)}_{s}}$,
\beqs
&& {\mbox{\large $K$}_{ {t_{i_1}}{t_{i_2}}\cdots {t_{i_{d-1}}}{t_{i_d}} }} \\
&& =\corr{ {{\cal O}^{(1)}_{t_{i_1}}}\,
{{\cal O}^{(1)}_{t_{i_2}}}\,\cdots \,
{{\cal O}^{(1)}_{t_{i_{d-1}}}}\,
{{\cal O}^{(1)}_{t_{i_d}}}} \\
&& ={{\left(
{\Kappa^{(1)}_{{t_{i_2}}}}
{\Kappa^{(2)}_{{t_{i_3}} }}
{\Kappa^{(3)}_{{t_{i_4}} }}
\cdots
{\Kappa^{(d-3)}_{{t_{i_{d-2}}} }}
{\Kappa^{(d-2)}_{{t_{i_{d-1}}} }}
{\Eta}^{(d-1)}
\right)}_{{t_{i_1}}{t_{i_d}}}} \\
&&=
\mbox{
\setlength{\unitlength}{1mm}
\begin{picture}(150,23)(0,22.5)
\put(20,15){\line(1,0){115}}
\put(40,15){\line(0,1){15}}
\put(55,15){\line(0,1){15}}
\put(70,15){\line(0,1){15}}
\put(85,15){\line(0,1){15}}
\put(100,15){\line(0,1){15}}
\put(115,15){\line(0,1){15}}
%%%%%%%
\put(37,33){\makebox(6,6){${\cal O}^{(1)}_{t_{i_2}}$}}
\put(52,33){\makebox(6,6){${\cal O}^{(1)}_{t_{i_3}}$}}
\put(67,33){\makebox(6,6){${\cal O}^{(1)}_{t_{i_4}}$}}
\put(82,33){\makebox(6,6){${\cal O}^{(1)}_{t_{i_{d-3}}}$}}
\put(97,33){\makebox(6,6){${\cal O}^{(1)}_{t_{i_{d-2}}}$}}
\put(112,33){\makebox(6,6){${\cal O}^{(1)}_{t_{i_{d-1}}}$}}
\put(9,12){\makebox(6,6){${\cal O}^{(1)}_{t_{i_1}}$}}
\put(140,12){\makebox(6,6){${\cal O}^{(1)}_{t_{i_d}}$}}
%%%%%%%%%%%%%
\put(20,15){\circle*{2}}
\put(40,30){\circle*{2}}
\put(55,30){\circle*{2}}
\put(70,30){\circle*{2}}
\put(85,30){\circle*{2}}
\put(100,30){\circle*{2}}
\put(115,30){\circle*{2}}
\put(135,15){\circle*{2}}
\put(72.5,19.5){\makebox(6,6){$\cdots$}}
\end{picture}
} \,\,\,,\\
&&
\eeqs
where the symbol $\{{t_{i_l}}\}$ takes $t$ or $s$.
Also the symbol $\Eta$ is the two-point function
(topological metric),
\beqs
&&{\Eta^{(d-1)}}:=\corr{{{\cal O}^{(d-1)}}
{{\cal O}^{(1)}}}
=\left(
\begin{array}{cc}
2 & 1 \\
1 & 0
\end{array}
\right) \,\,\,.
\eeqs
Because of the relation (\ref{eqn:kaseki}),
the above d-point functions are independent of the
positions of the inserted external fields ${{\cal O}^{(1)}_{t_i}}$
and we may write the d-point couplings having $n$ ${{\cal O}^{(1)}_{s}}$'s
and ${(d-n)}$ ${{\cal O}^{(1)}_{t}}$'s as $\mbox{\large ${K}$}_{d-n,n}$
in an abbreviated form. Then these couplings are represented in a matrix form,
\beqs
%&& \mbox{\Large $K$}_{d-n,n}\\
%&& =
&&{{\left(
{\Kappa^{(1)}_{{t}}}
{\Kappa^{(2)}_{{t} }}\cdots
{\Kappa^{(d-3-l)}_{{t} }}
{\Kappa^{(d-2-l)}_{{t} }}
{\Kappa^{(d-1-l)}_{{s} }}
{\Kappa^{(d-l)}_{{s} }}
\cdots
{\Kappa^{(d-3)}_{{s} }}
{\Kappa^{(d-2)}_{{s} }}
{\Eta}^{(d-1)}
\right)}_{{t_{i_1}}{t_{i_d}}} } \\
&&=
\mbox{
\setlength{\unitlength}{1mm}
\begin{picture}(160,33)(0,22.5)
\put(20,15){\line(1,0){120}}
\put(35,15){\line(0,1){20}}
\put(45,15){\line(0,1){20}}
\put(55,15){\line(0,1){20}}
\put(65,15){\line(0,1){20}}
\put(75,15){\line(0,1){20}}
\put(85,15){\line(0,1){20}}
\put(95,15){\line(0,1){20}}
\put(105,15){\line(0,1){20}}
\put(115,15){\line(0,1){20}}
\put(125,15){\line(0,1){20}}
%%%%%%%%%%%%%%%%%%%%%%
\put(67,22){\makebox(6,6){$\cdots$}}
\put(107,22){\makebox(6,6){$\cdots$}}
%%%%%%%%%%%%%%%%%%%%
\put(32,38){\makebox(6,6){${{\cal O}_t}$}}
\put(42,38){\makebox(6,6){${{\cal O}_t}$}}
\put(52,38){\makebox(6,6){${{\cal O}_t}$}}
\put(62,38){\makebox(6,6){${{\cal O}_t}$}}
\put(72,38){\makebox(6,6){${{\cal O}_t}$}}
\put(82,38){\makebox(6,6){${{\cal O}_t}$}}
\put(92,38){\makebox(6,6){${{\cal O}_s}$}}
\put(102,38){\makebox(6,6){${{\cal O}_s}$}}
\put(112,38){\makebox(6,6){${{\cal O}_s}$}}
\put(122,38){\makebox(6,6){${{\cal O}_s}$}}
%%%%%%%%%%%%%%%%%%%%%%
\put(54,48){\makebox(12,6){$d-2-l$}}
\put(104,48){\makebox(12,6){$l$}}
%%%%%%%%%%%%%%%%%%%%%
\put(60,42){\oval(58,4)[t]}
\put(110,42){\oval(38,4)[t]}
%%%%%%%%%%%%%%%%%%%%%%%
\put(20,15){\circle*{2}}
\put(140,15){\circle*{2}}
\put(35,35){\circle*{2}}
\put(45,35){\circle*{2}}
\put(55,35){\circle*{2}}
\put(65,35){\circle*{2}}
\put(75,35){\circle*{2}}
\put(85,35){\circle*{2}}
\put(95,35){\circle*{2}}
\put(105,35){\circle*{2}}
\put(115,35){\circle*{2}}
\put(125,35){\circle*{2}}
\end{picture}
}\\
&& \\
&&\mbox{} \\
%&&\mbox{}\\
&&=\left(
\begin{array}{cc}
\mbox{\large $K$}_{d-l,l} & \mbox{\large $K$}_{d-l-1,l+1} \\
\mbox{\large $K$}_{d-l-1,l+1} & \mbox{\large $K$}_{d-l-2,l+2}
\end{array}
\right) \,\,\,.
\eeqs
We find these d-point couplings,
\beqs
&& \mbox{\large $K$}_{d,0}
=2{(d+1)}\\
&&\hspace{1.5cm}
+2{(d+1)} \cdot \Big\{{2\cdot {(d+1)}^{(d+1)} -(d+2)\cdot
 {(d+1)}!-d \cdot {(d+1)}!
\left({\sum^{d+1}_{m=2} \frac{d+1}{m}}\right)}\Big\} q_1
+\mbox{\large $O$}({q^2}) \,\,\,,\\
&& \mbox{\large $K$}_{d-1,1}
=(d+1)\\
&&\hspace{1.5cm}
+(d+1) \cdot \Big\{{{(d+1)}^{(d+1)} -2\cdot {(d+1)}!-(d-1)\cdot {(d+1)}!
\left({\sum^{d+1}_{m=2}
 \frac{d+1}{m}}\right)}\Big\} q_1
+\mbox{\large $O$}({q^2}) \,\,\,,\\
&& \mbox{\large $K$}_{d-2,2}
=0+\mbox{\large $O$}({q^2}) \,\,\,,\\
&& \mbox{\large $K$}_{d-3,3}
=(d+1) {q_2} +\mbox{\large $O$}({q^2}) \,\,\,,\\
&& \mbox{\large $K$}_{d-n,n}
=0+\mbox{\large $O$}({q^2}) \,\,\,,\,\,(n \geq 4)\,\,\,.
\eeqs

In this article, we treated the one-parameter models and the
two-parameter model concretely, but
the method developed in this article is not restricted to these cases only.
%%%%%%%%%%%ddddddddddddddd

\section{Conclusion}

\pr
In this article, we have investigated some properties of the higher dimensional
Calabi-Yau manifolds subject to the assumption of the existence of the
mirror symmetries. We extend the method of calculating the
three point functions for
the one-parameter families of $d$-folds and a
two-parameter family of $d$-fold for Calabi-Yau cases.
The recipe developed here can be applied to more general cases (for instance
the complete intersection Calabi-Yau $d$-folds in the toric cases).
Explicit forms of the homology cycles are not available, but
the monodromy properties of the period matrix illustrate the correctness
 of our Ansatz. We used the mirror conjecture and
our results should be verified by the mathematical methods in enumerative
geometry {\cite{K}}.

For the general {\kae} manifolds $M$, the virtual dimension
of the A(M)-model has a term depending on the first Chern class and
the degree of maps.
Because of the existence of this term,
the degree of the observables are defined modulo
${{c_1}(M)}$.
(There exists each topological selection rule when one fixes the
degree of maps). Also from the point of view of the deformation of the
topological field theories, we perturb the topological theories by
adding only operators associated with the {\kae} forms of M in our
cases. In these situations, the charge conservation in each fusion of
operators can be discussed almost classically. Only difference
between the A(M)-model with ${c_1}\mbox{(M)}{=0} $ and the one with
${c_1}\mbox{(M)}{\not=}0 $ is that the former has nilpotent structures
of operators ${{\cal O}^{(1)}}{{\cal O}^{(d)}}=0 $ but the latter does
not have these properties. Also the fusion couplings of operators in
the former cases have contributions from all degrees of maps because
the virtual dimension is independent of the degree of maps. That is
to say, the three-point couplings in the Calabi-Yau cases are infinite
series with respect to indeterminates ${{q_l}:={e^{2\pi i {t_l}}}}
\,\,({l=1,2,\cdots \,,\dim {H^{1,1}}\mbox{(M)}}) $ associated with a
set of {\kae} coordinates $\{{t_l}\}$.

\section*{Acknowledgement}
\pr
I would like to express my sincere gratitude to
Prof.~T.~Eguchi for guidance
and kindful encouragement throughout my graduate course.
I also thank Prof.~S.~Hosono and Dr.~K.~Hori
for useful discussions and comments.


\newpage
\begin{thebibliography}{99}

\bibitem{W} E.~Witten,
{\it Mirror Manifolds and Topological Field Theory}, in
{\it Essays on Mirror Manifolds}, ed. S.-T.~Yau,
(Int. Press, Hong Kong, 1992), pp.120-180.
\bibitem{WIT} E.~Witten, Commun. Math. Phys. {\bf 118} (1988) 441.
\bibitem{EY} T.~Eguchi and S.-K.~Yang, Mod. Phys. Lett. {\bf A5}
(1990) 1693.
\bibitem{DG} J.~Distler and B.~Greene, Nucl.~Phys. {\bf B309} (1988) 295.
\bibitem{CDGP} P.~Candelas, X.~de la Ossa, P.~Green and L.~Parkes,
Phys.~Lett. {\bf 258B} (1991) 118; Nucl.~Phys. {\bf B359} (1991) 21.
\bibitem{AM} P.~Aspinwall, D.~Morrison,
Commun.~Math.~Phys. {\bf 151} (1993) 245.
\bibitem{GP} B.~Greene and M.~Plesser, Nucl.~Phys. {\bf B338} (1990) 15.
\bibitem{CLS} P.~Candelas, M.~Lynker and R.~Schimmrigk,
Nucl.~Phys. {\bf B341} (1990) 383.
\bibitem{Y} {\it Essays on Mirror Manifolds}, ed. S.-T.~Yau,
(Int. Press, Hong Kong, 1992).
\bibitem{KT} A.~Klemm and S.~Theisen, Nucl.~Phys.
{\bf B389} (1993) 153.
\bibitem{F} A.~Font, Nucl.~Phys. {\bf B391} (1993) 358.
\bibitem{KT2} A.~Klemm and S.~Theisen,
Mod.~Phys.~Lett. {\bf A9} (1994) 1807.
%"Mirror Maps and
%Instanton Sums for Complete Intersection in Weighted projective
%Space", preprint LMU-TPW 93-08.
\bibitem{CDFKM} P.~Candelas, X.~de la Ossa, A.~Font, S.~Katz
and D.~Morrison, Nucl.~Phys. {\bf B416} (1994) 481.
\bibitem{CFKM} P.~Candelas, A.~Font, S.~Katz
and D.~Morrison, Nucl.~Phys. {\bf B429} (1994) 626.
\bibitem{BCDFHJQ} P.~Berglund, P.~Candelas, X.~de~la~Ossa, A.~Font,
T.~H{\"u}bsch, D.~Jan{\v{c}}i{\'c} and F.~Quevedo,
Nucl.~Phys. {\bf B419} (1994) 352.
\bibitem{LSW} W.~Lerche, D.~Smit and N.~Warner,
Nucl.~Phys. {\bf B372} (1992) 87.
\bibitem{HKT} S.~Hosono, A.~Klemm and S.~Theisen,
"Lectures on Mirror Symmetry", HUTMP-94/01, LMU-TPW-94-02.\\
S.~Hosono, A.~Klemm, S.~Theisen, and S.-T.~Yau,
Commun.~Math.~Phys. {\bf 167} (1995) 301;
Nucl.~Phys. {\bf B433} (1995) 501.
\bibitem{BV} V.~Batyrev and D.~van Straten,
"Generalized Hypergeometric Functions and Rational Curves
on Calabi-Yau Complete Intersections in Toric Varieties",
Essen preprint, alg-geom/9307010.
\bibitem{NS} M.~Nagura and K.~Sugiyama,
Int.~J.~Mod.~Phys. {\bf A10} (1995) 233.
\bibitem{GMP} B.~Greene, D.~Morrison and M.~Plesser,
"Mirror Manifolds in Higher Dimension", CLNS-93/1253, IASSNS-HEP-94/2,
YCTP-P31-92.
%%%%%%%%%%%%%%%%
\bibitem{JN} M.~Jinzenji and M.~Nagura,
"Mirror Symmetry and An Exact Calculation of $N-2$ Point
Correlation Function on Calabi-Yau Manifold embedded in
${{CP}^{N-1}}$", preprint UT-680.
\bibitem{G} P.~Griffiths, Ann.~Math. {\bf 90} (1969) 460, 469.
\bibitem{BG} R.~Bryant and P.~Griffiths,
Progress in mathematics {\bf 36} (Birkh{\"a}user, Boston, 1983), p.~77.
\bibitem{K}
%M.~Kontsevich,
%"Enumeration of Rational Curves
%via Torus Actions", preprint.\\
M.~Kontsevich and Y.~I.~Manin,
%"Gromov-Witten classes, Quantum
%Cohomology, and Enumerative Geometry", preprint.
Commun.~Math.~Phys. {\bf 164} (1994) 525.

\end{thebibliography}
\end{document}